\documentclass[subscriptcorrection,upint,varvw,barcolor=black,mathalfa=cal=euler,balance,hyphenate,french,pdf-a,nolists]{asmejour} %
\usepackage{makecell}
\usepackage{multirow} 
\hypersetup{ pdfauthor={Weixuan Li, Sharun Kuhar, Jung-Hee Seo, Rajat Mittal},
	pdftitle={Modeling the Effect of Sleeve Gastrectomy on Gastric Digestion in the Stomach: Insights from Multiphase Flow Modeling},                  	
	pdfkeywords={Computational Fluid Dynamics, Multiphase Flow, Surgery, Laparoscopic Sleeve Gastrectomy}
	   }

\JourName{Biomechanical Engineering}

\begin{document}


\SetAuthorBlock{Weixuan Li}{Department of Mechanical Engineering,\\
   Johns Hopkins University,\\
   3400 N Charles Street,\\
   Baltimore, MD 21218 \\
   email: wli142@jh.edu} 

\SetAuthorBlock{Sharun Kuhar}{Department of Mechanical Engineering,\\
   Johns Hopkins University,\\
   3400 N Charles Street,\\
   Baltimore, MD 21218 \\
   email: skuhar1@jhu.edu}

\SetAuthorBlock{Jung-Hee Seo}{Department of Mechanical Engineering,\\
   Johns Hopkins University,\\
   3400 N Charles Street,\\
   Baltimore, MD 21218 \\
   email: jhseo@jhu.edu}
   

\SetAuthorBlock{Rajat Mittal\CorrespondingAuthor}{Department of Mechanical Engineering,\\
   Johns Hopkins University,\\
   3400 N Charles Street,\\
   Baltimore, MD 21218 \\
   email: mittal@jhu.edu} 

\title{Modeling the Effect of Sleeve Gastrectomy on Gastric Digestion in the Stomach: Insights from Multiphase Flow Modeling}

\keywords{Computational Fluid Dynamics, Multiphase Flow, Surgery,
Laparoscopic Sleeve Gastrectomy}

\begin{abstract}
The geometry and motility of the stomach play a critical role in the digestion of ingested liquid meals. Sleeve gastrectomy, a common type of bariatric surgery used to reduce the size of the stomach, significantly alters the stomach’s anatomy and motility, which impacts gastric emptying and digestion. In this study, we use an imaging data-based computational model, \textit{StomachSim}, to investigate the consequences of sleeve gastrectomy. The pre-operative stomach anatomy was derived from imaging data and the post-sleeve gastrectomy shapes were generated for different resection volumes. We investigate the effect of sleeve sizes and motility patterns on gastric mixing and emptying. Simulations were conducted using an immersed-boundary flow solver, modeling a liquid meal to analyze changes in gastric content mixing and emptying rates. The results reveal that different degrees of volume reduction and impaired gastric motility have complex effects on stomach's mixing and emptying functions, which are important factors in gastric health of the patient. These findings provide insights into the biomechanical effects of sleeve gastrectomy on gastric digestion and emptying functions, highlighting the potential of computational models to inform surgical planning and post-operative management. 
\end{abstract} 

\date{Version \versionno, \today}

\maketitle 



\section{Introduction}
Since 1980, the global prevalence of people who are considered overweight or obese has doubled, and by 2022, 1 in 8 people worldwide were living with obesity. In the same year, over 2.5 billion adults were overweight, including 890 million who were classified as obese~\cite{who2024obesity}. Sleeve gastrectomy (SG), specifically laparoscopic sleeve gastrectomy (LSG), is a weight loss surgery designed to help extremely obese patients reduce their weight~\cite{yehoshua2008laparoscopic}. The LSG procedure has gained popularity and become a favored choice among bariatric surgeons~\cite{iannelli2008laparoscopic}. Laparoscopic sleeve gastrectomy, which involves removing a significant portion of the stomach via a small abdominal incision, not only reduces gastric volume but also alters key stomach functions, including compliance, adaptive reflexes, and regulation of emptying speed~\cite{sista2017effect}. It also has a significant effect on stomach peristalsis, which is an important factor in the gastric phase of digestion and which refers to the process by which the stomach stores, grinds, and delivers food. Gastric motility and electrical activity are significantly impaired due to the resection of most of the fundus and the gastric pacemaker located along the greater curvature~\cite{csendes2016changes}. Baumann \emph{et al.} conducted an experiment on gastric motility after sleeve gastrectomy and found that while antral motility was preserved in these patients, the sleeve exhibited no recognizable peristalsis in three of the five patients, and only uncoordinated or passive motion in the remaining two. These factors all have a significant impact on gastric digestion and emptying.

Experiments have been conducted to study the effects of sleeve gastrectomy on the stomach function. Garay \emph{et al.} evaluated the impact of antrum size on gastric emptying and weight loss outcomes after LSG~\cite{garay2018influence}.  Braghetto \emph{et al.} evaluated the impact of LSG on gastric emptying in obese patients compared to normal subjects, revealing significantly accelerated emptying of liquids and solids after surgery~\cite{braghetto2009scintigraphic}. Some studies present a delayed emptying rate~\cite{wickremasinghe2024modified}, while most investigations indicate a rapid emptying ~\cite{sista2017effect, braghetto2009scintigraphic, samuel2016effect, kara54, kandeel2015comparative}. However, these studies only evaluated the emptying rate by measuring the half emptying time $T_{1/2}$ without providing detailed information on the emptying process~\cite{sista2017effect, kandeel2015comparative}. 

Computational methods offer significant advantages in biomechanical investigations by reducing the need for animal experiments, enhancing cost and time efficiency, and providing detailed insights into biological structures and surgical procedures that are difficult to achieve through traditional experimental methods~\cite{carniel2020computational}. There are relatively few studies that have attempted to investigate sleeve gastrectomy using computational models. Toniolo \emph{et al.} developed computational models to analyze biomechanical changes in the stomach before and after LSG, focusing on the effects of volumetric reduction, gastric wall stiffness, and elongation strain distribution under different pressures~\cite{toniolo2021computational, toniolo2022patient}. Computational studies for this surgery primarily focus on the volume-pressure relationship and the properties of the stomach wall, with limited attention to the interactions involving gastric contents and emptying functions. 

In this study, we perform multiphase flow simulations on a full stomach model derived from Magnetic Resonance Imaging (MRI) data and post-operative models with reduced gastric volume and motility to explore the effects of LSG on stomach digestion and emptying functions, focusing on the emptying rate and the mixing of gastric contents. The goal is to demonstrate the potential for such models to provide insights into the gastric mixing and emptying processes, offering valuable clinical information related to this bariatric surgery.

\section{Methods} 
%
\subsection{Stomach Model}
\begin{figure}[h] 
    \centering
    \includegraphics[width=0.46\textwidth]{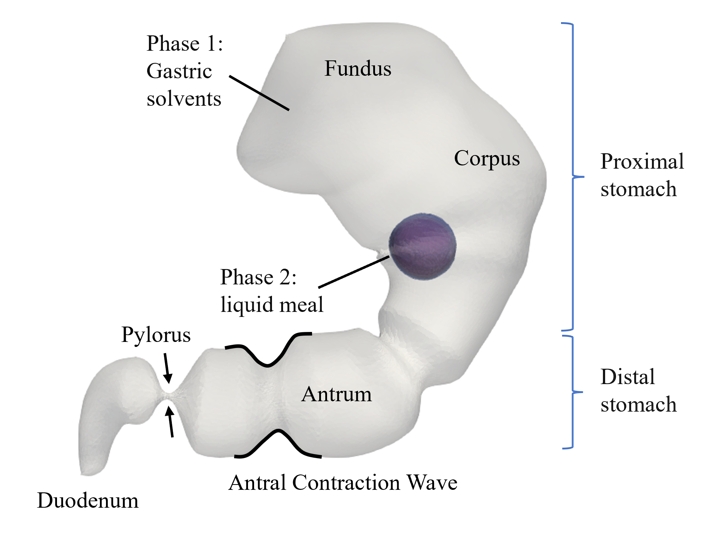}
    \caption{Schematic configuration of human stomach and initial condition of the pre-operative stomach model.} 
    \label{stomach_initial}
\end{figure}

The pre-operative gastrectomy stomach geometry is a model based on the model used in our previous work ~\cite{kuhar2022effect}. This stomach model was segmented from the Virtual Population Library (VPL)~\cite{gosselin2014development} for an adult male ("Duke"). However, the VPL geometry features a tubular antrum which is characteristic of an empty or a low food volume stomach. We modified the VPL model to make it resemble a postprandial stomach based on publicly available MRI data such as the study by Lu \emph{et al.}~\cite{lu2022automatic} and the website of Motilent (London, England). The structure of the human stomach with a full bolus inside is shown in Fig.~\ref{stomach_initial}. The stomach has three regions: the fundus, corpus, and antrum. The fundus and corpus are located in the proximal stomach and primarily function as storage areas for food. The antrum, located in the distal stomach, is the main site for mixing, grinding, and digestion of gastric contents. Gastric motility is driven by antral contraction waves (ACWs) initiated by a pacemaker located on the greater curvature of the corpus. These waves progressively increase in strength toward the antrum, playing a crucial role in gastric motility. ACWs occur every period (20 seconds). During each period, the pylorus—a 2 mm orifice—remains open for the first 7 seconds and closed thereafter, allowing gastric contents to empty.

We consider a full bolus of a high-viscosity liquid meal like honey. Two-phase flows are simulated inside the stomach. Phase 1 represents the gastric solvent, assumed to have the properties of water, with a density of $\rho = 1000\, \text{kg/m}^3$ and a viscosity of $\mu = 0.001\, \text{Pa}\cdot\text{s}$. Phase 2 represents the liquid meal ingested into the stomach, with a density of $\rho = 1020\, \text{kg/m}^3$, and a viscosity of $\mu = 0.1\, \text{Pa}\cdot\text{s}$ which is 100 times that of Phase 1. Initially, the stomach is filled with gastric liquid, and a 5\,mL drop of bolus is placed in the proximal stomach to represent the liquid meal swallowed through the esophagus. The initial condition of the pre-operative stomach model is shown in Fig.~\ref{stomach_initial}. In the other cases, the initial conditions are the same, with the drop initially located at the same position.

\subsection{Modeling Surgery}
\begin{figure}[h]
    \centering
    \begin{tabular}{cc}
        \begin{subfigure}[b]{0.23\textwidth} 
            \centering
            \includegraphics[height=4.5cm]{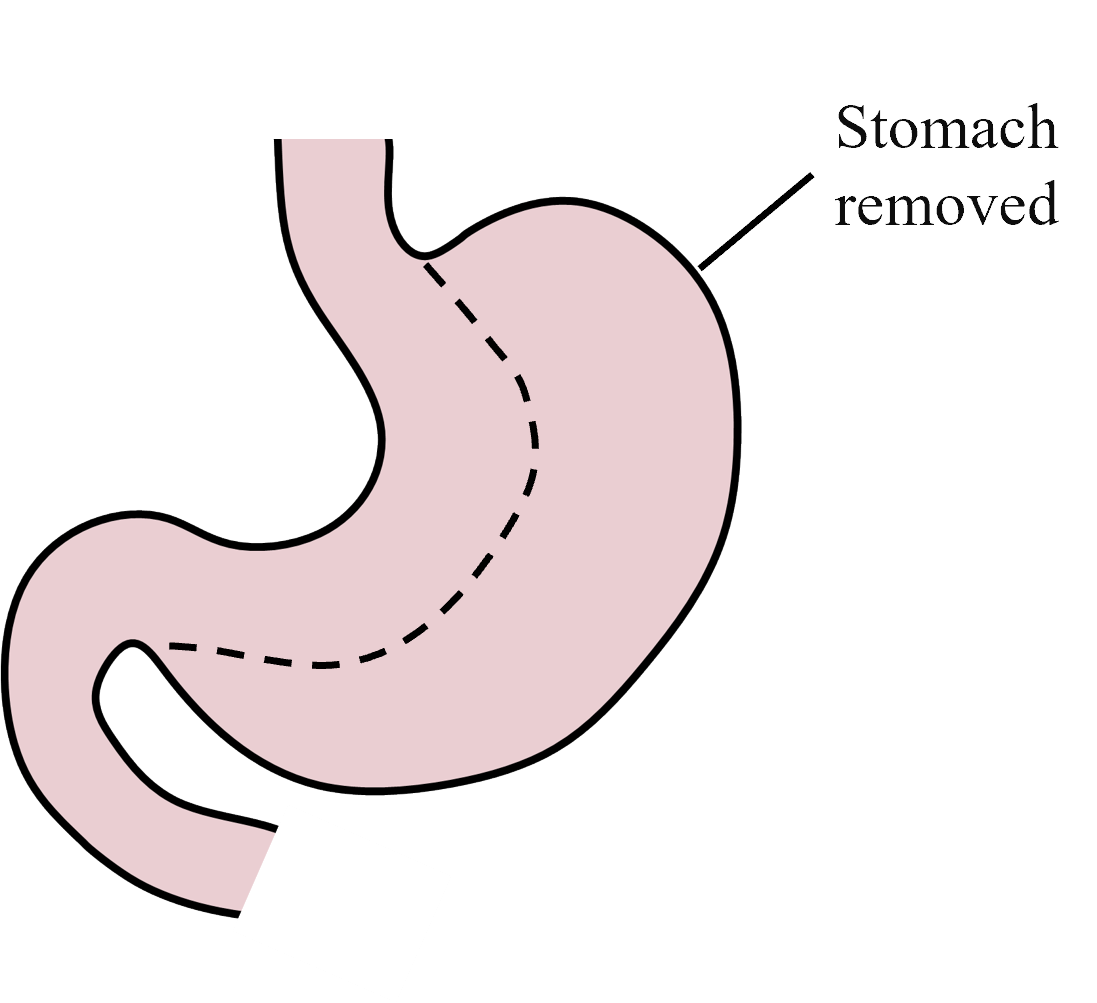} 
            \caption{Pre-operative stomach.}
            \label{lsg1}
        \end{subfigure} &
        \begin{subfigure}[b]{0.23\textwidth}
            \centering
            \includegraphics[height=4.5cm]{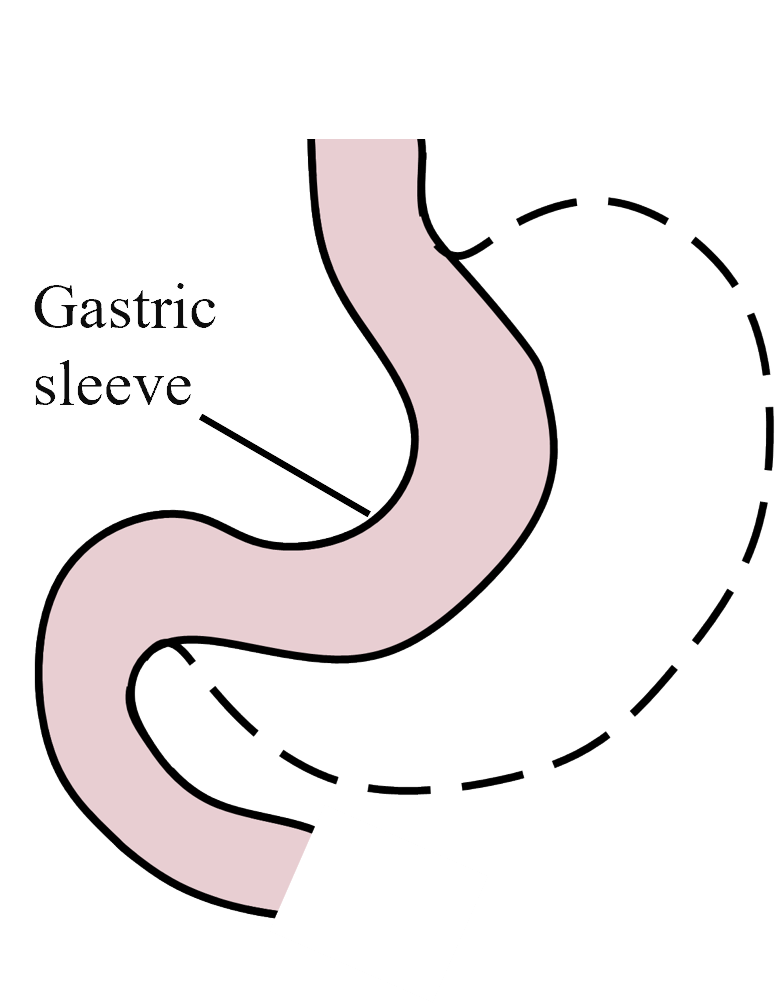} 
            \caption{Post-operative stomach.}
            \label{lsg2}
        \end{subfigure}
    \end{tabular}

    \caption{Schematic configuration of laparoscopic sleeve gastrectomy procedure}
    \label{lsg}
\end{figure}

Laparoscopic sleeve gastrectomy is a permanent surgical procedure conducted in a hospital under general anesthesia. The schematic configuration of the surgery procedure is shown in Fig.~\ref{lsg}. During the operation, the surgeon makes approximately five small incisions in the abdomen. A thin, long scope with a tiny camera at the end is utilized to guide the surgery. Through these incisions, instruments are inserted to remove a significant portion of the stomach, specifically the outward-curving section known as the fundus and corpus. After its removal, the remaining part of the stomach is shaped into a tube resembling a banana or a shirt sleeve—hence the name "sleeve gastrectomy"~\cite{johns2024laparoscopic}. With a much smaller stomach, patients will feel full more quickly during meals and will eat less.

\begin{figure}[h]
    \centering
    \begin{subfigure}[b]{0.23\textwidth} 
        \centering
        \includegraphics[width=\textwidth]{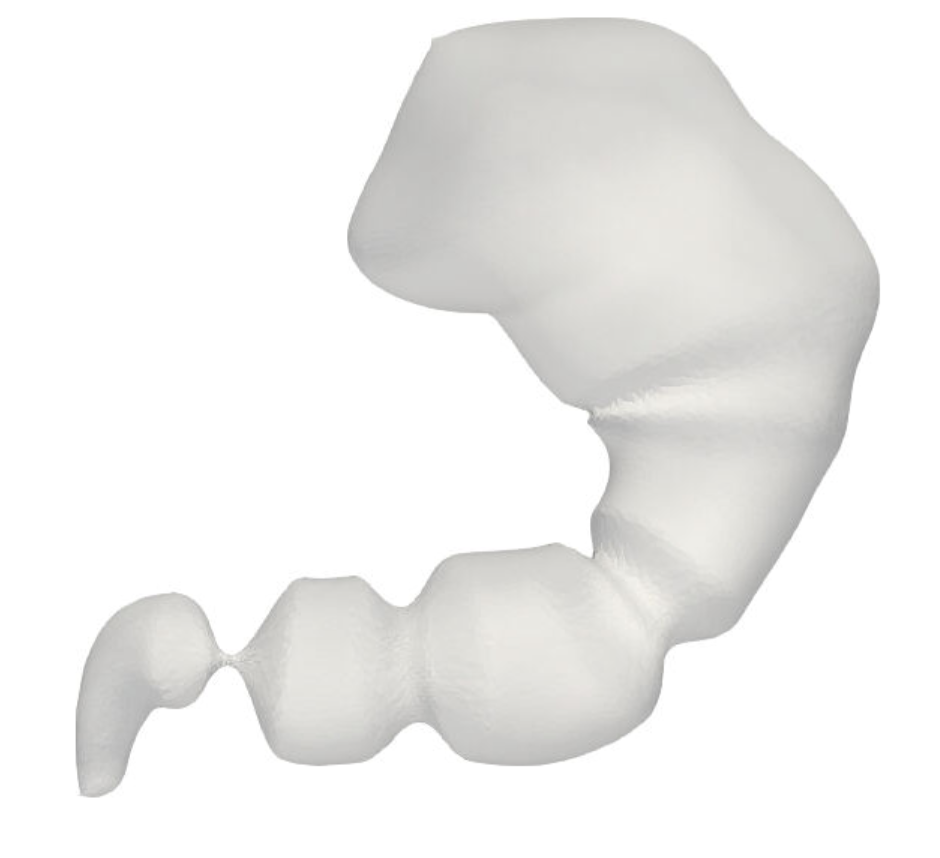}
        \caption{Pre-operative}    
        \label{entire}
    \end{subfigure}
    
    \vspace{0.5em} 
    \begin{tabular}{cc}
        \begin{subfigure}[b]{0.23\textwidth} 
            \centering
            \includegraphics[height=\textwidth]{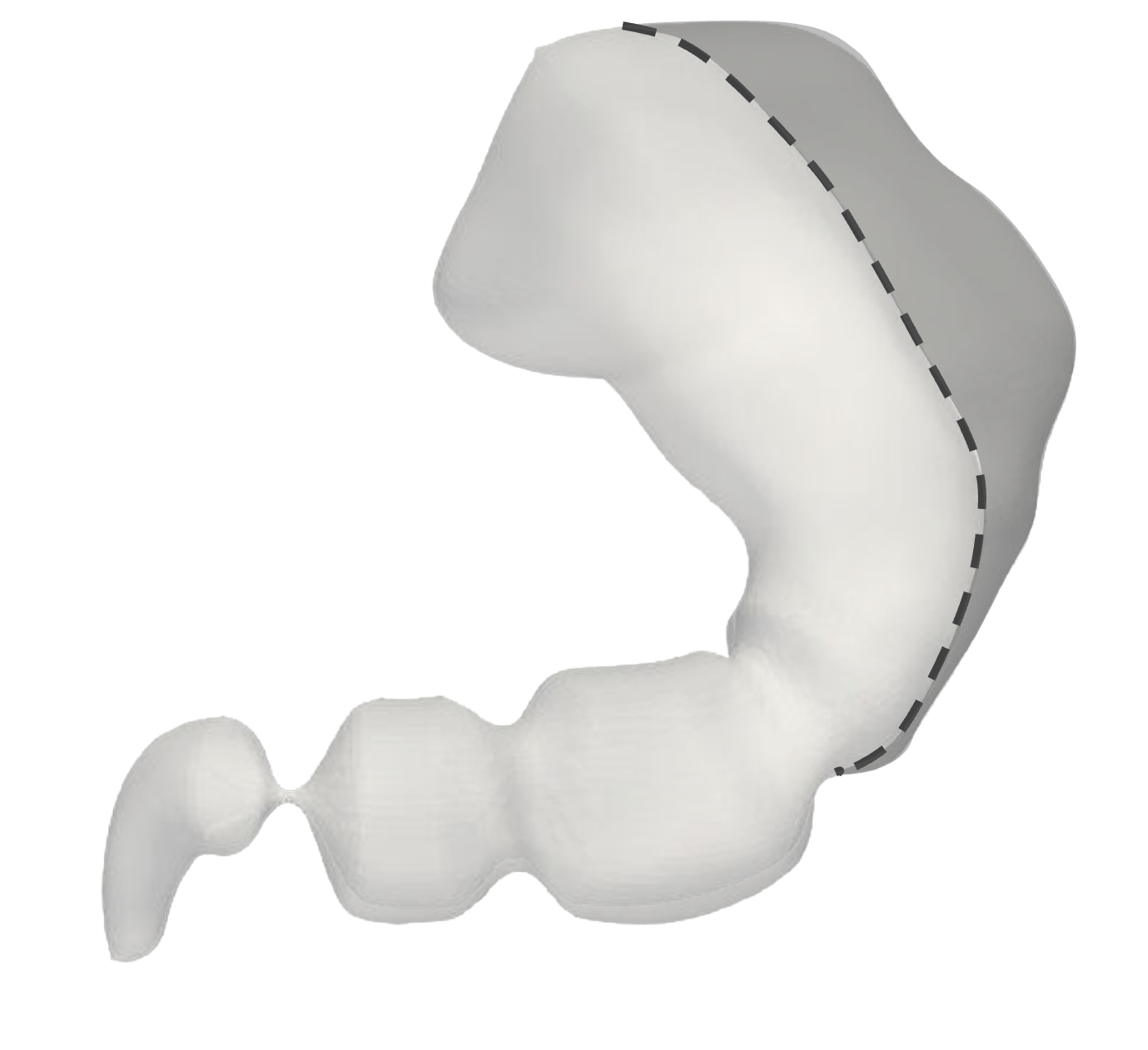} 
            \caption{Post-operative, moderate sleeve (70\% of pre-op. volume)}
            \label{large_sleeve}
        \end{subfigure} &
        \begin{subfigure}[b]{0.23\textwidth}
            \centering
            \includegraphics[height=\textwidth]{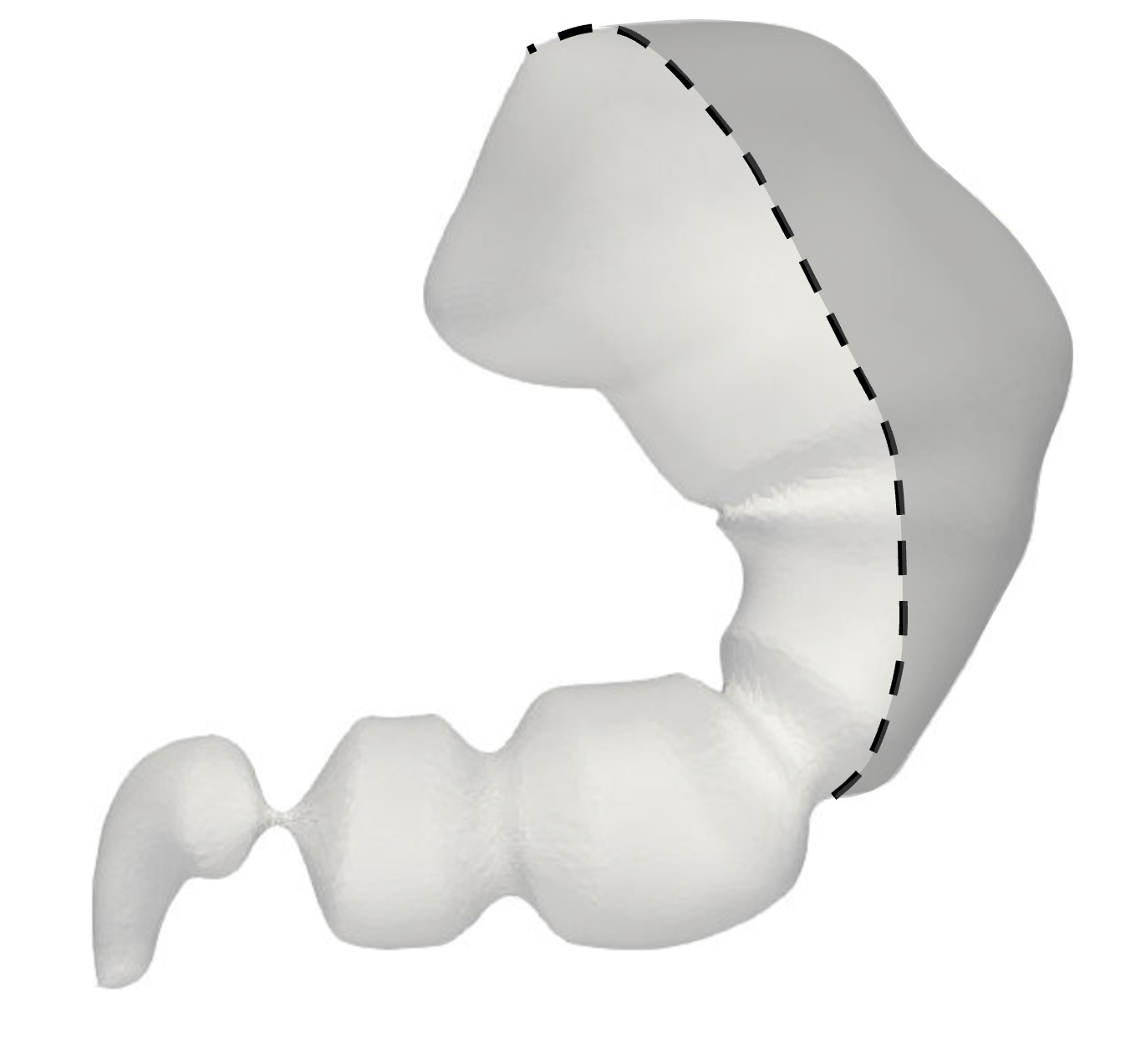} 
            \caption{Post-operative, extreme sleeve (45\% of pre-op. volume)}
            \label{small_sleeve}
        \end{subfigure}
    \end{tabular}

    \caption{Pre- and post-operative stomach models. Post-operative models illustrate the effects of varying resection volumes on stomach geometry.}
    \label{geometry}
\end{figure}

Computational post-operative models were constructed using Blender (v4.1)~\cite{blender41} and are shown in Fig.~\ref{geometry}. During LSG, a significant portion of the fundus and a small section of the antrum are resected~\cite{sista2017effect}. The extent of stomach resection in actual surgeries varies from less than 50\% to more than 90\%~\cite{toniolo2022patient}. To generalize the simulations, we created two post-operative stomach models with varying resected volumes: the moderate sleeve stomach (Fig.~\ref{large_sleeve}) retains 70\% of the pre-operative stomach volume, while the extreme sleeve stomach (Fig.~\ref{small_sleeve}) retains 45\% of the pre-operative stomach volume.


The motility patterns are dynamically adjusted based on the activity of the pacemaker. In the absence of the pacemaker, gastric motility in the proximal stomach experiences varying degrees of reduction. In the study by Baumann \emph{et al.}, motility in the sleeve tube above the antrum was impaired significantly. Among the five patients studied by Baumann \emph{et al.}, three exhibited no peristalsis in the sleeve tube, while two showed partial contraction patterns, which differed between the patients. However, antral motility was preserved~\cite{baumann2011time}. Additionally, the resection end points varied from $2\, \text{cm}$ to over $9\, \text{cm}$ distal to the pylorus between different surgeries~\cite{csendes2016changes}.

\begin{figure}[h]
    \centering
    \begin{tabular}{cc}
        \begin{subfigure}[b]{0.23\textwidth} 
            \centering
            \includegraphics[height=4cm]{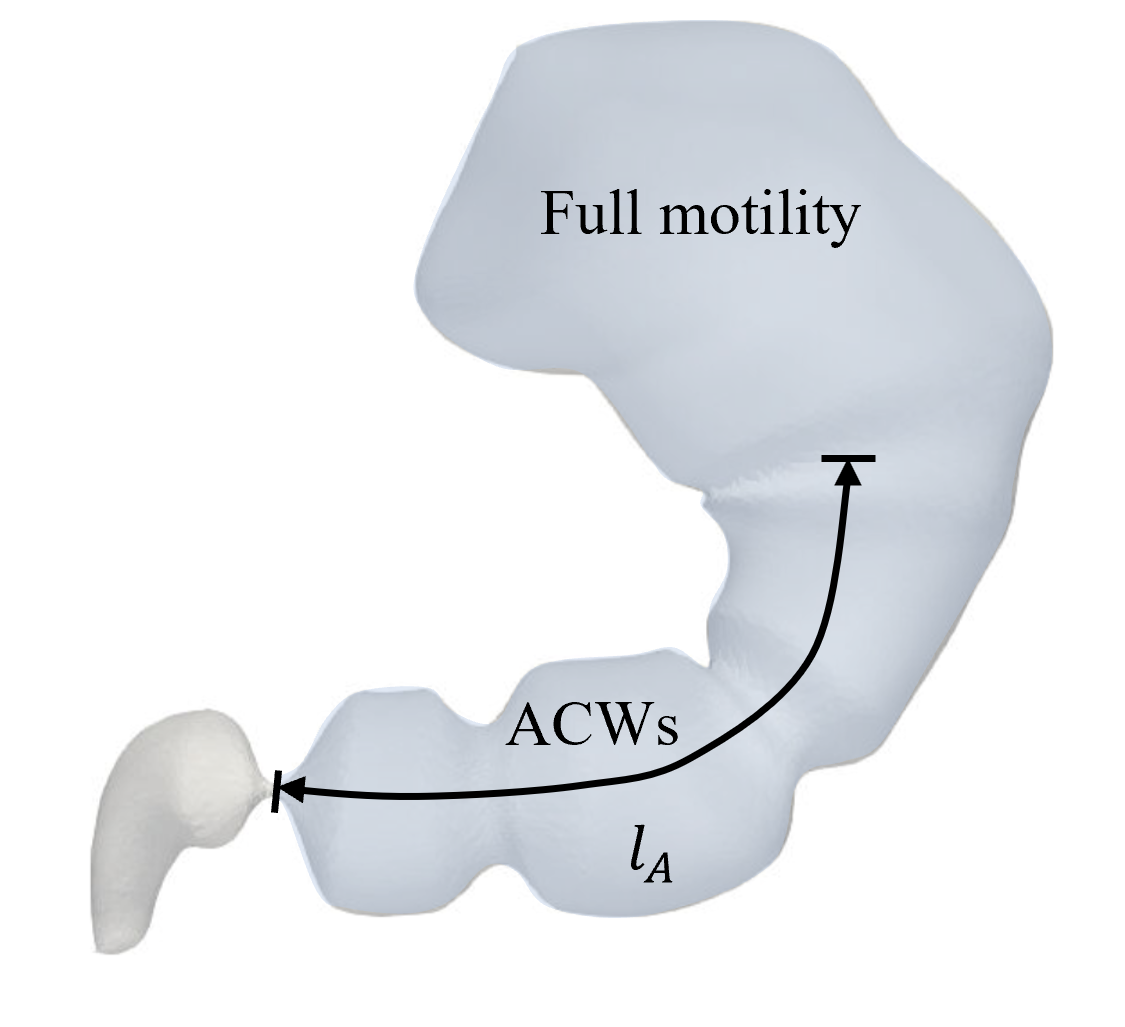} 
            \caption{Pre-operative stomach model with full motility. The centerline length of the region where ACWs exist is $l_A = 8\, \text{cm}$.}
            \label{pre-op}
        \end{subfigure} &
        \begin{subfigure}[b]{0.23\textwidth}
            \centering
            \includegraphics[height=4cm]{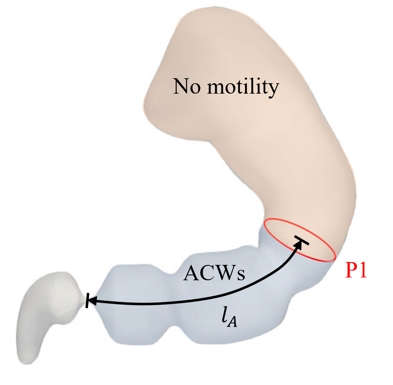} 
            \caption{Moderate sleeve stomach model. Resection ends at P1 ($6\, \text{cm}$ from the pylorus) with no motility in the sleeve tube.}
            \label{ll}
        \end{subfigure}
    \end{tabular}
    
    \vspace{1em} 
    \begin{tabular}{cc}
        \begin{subfigure}[b]{0.23\textwidth} 
            \centering
            \includegraphics[height=4cm]{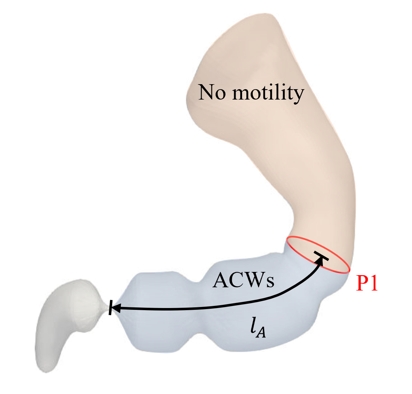} 
            \caption{Extreme sleeve stomach model. Resection also ends at P1 ($6\, \text{cm}$ from the pylorus) with no motility in the sleeve tube.}
            \label{sl}
        \end{subfigure} &
        \begin{subfigure}[b]{0.23\textwidth}
            \centering
            \includegraphics[height=3.9cm]{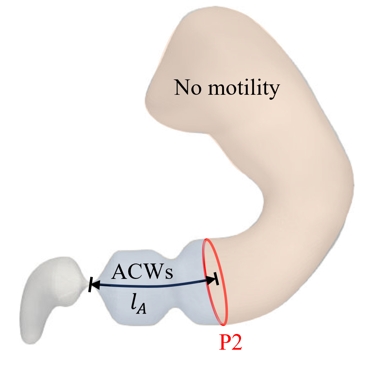} 
            \caption{Moderate sleeve model with reduced motility. Resection ends at P2 ($2\, \text{cm}$ from pylorus) without sleeve tube motility.}
            \label{ll_reduced}
        \end{subfigure}
    \end{tabular}
    
    \caption{Illustration of four cases representing the pre-operative stomach and potential outcomes of sleeve gastrectomy.}
    \label{motility}
\end{figure}

Based on these previous studies on gastric motility after LSG, we designed four distinct cases representing stomach models with different volumes and motility patterns to simulate the pre-operative stomach and various potential outcomes of sleeve gastrectomy, as illustrated in Fig.~\ref{motility}. The information of four cases and corresponding parameters used are summarized in Table~\ref{four_cases}. The first case represents the pre-operative stomach with the full volume and normal motility. The length of the centerline in the region where the ACWs occur is $l_A = 8\, \text{cm}$. This model is referred to as the \emph{pre-operative stomach model} in subsequent discussions (Fig.~\ref{pre-op}). In the second case, we consider a moderate sleeve stomach model without motility in the sleeve tube, where the resection ends at point P1, located $6\, \text{cm}$ from the pylorus (i.e. $l_A = 6 cm$). This model is referred to as the \emph{moderate sleeve stomach model} (Fig.~\ref{ll}). The third case examines an extreme sleeve stomach, where the resection also ends at point P1 ($l_A = 6 , \text{cm}$), and the same motility patterns as the second case are employed to further analyze the effects of reduced fundus volume. This model is termed the \emph{extreme sleeve stomach model} (Fig.~\ref{sl}). In the fourth case, we examine a moderate sleeve stomach without motility in the sleeve tube, but with the resection ending at point P2, located $2\, \text{cm}$ from the pylorus (i.e. $l_A = 2 cm$). This model is referred to as the \emph{moderate sleeve model with reduced motility} or \emph{reduced motility model} (Fig.~\ref{ll_reduced}). 

A model for the imposition of fundic pressure was implemented to mimic the effect of fundic tone on the gastric contents. In the stomach, the fundus initially expands to accommodate the arrival of food in the stomach but then muscular contractions of the fundus impose a pressure on the fluid contents that plays a key role in gastric emptying. In the current computational model, the duodenal outlet pressure is set to a reference value of zero. At the fundic inlet boundary, the fundic contraction is modeled via  a fundic "piston" at the fundus opening that is driven inwards by a constant external pressure $p_o$. This condition can be expressed in terms of the rate-of-change in time of the normal flow velocity ($u_n$) at the piston:
        \begin{equation}\label{eq:pistbc}
            \frac{d{u}_n}{dt} = \frac{p_o-\bar{p}_A}{m_p}
        \end{equation}
where $\bar{p}_A$ is the average pressure on the fluid side of the piston and where $m_p$ is the mass per unit area of the piston. The above condition can be discretized using an explicit scheme to provide a value of velocity at the boundary. For pressure, we specify a zero gradient at the fundus opening, i.e. $\partial p / \partial n=0$. Further details regarding this boundary condition can be found in \cite{kuhar2024silico}.

We note that in the model of fundic tone above, $p_o$, which represents the pressure exerted by the fundus, affects the flux at the fundic inlet, and consequently at the duodenal outlet. The fundic pressure $p_o$ can therefore be prescribed to not just achieve the desired emptying rate into the duodenum, but also to model the effect of sleeve gastrectomy on the gastric tone. 
According to a previous study on the pressure-volume relationship after sleeve gastrectomy conducted by Toniolo \emph{et al.}~\cite{toniolo2021computational}, the intragastric pressure in post-operative models that retain approximately 40\% of the original stomach volume, increases to 1.67 times that of the pre-operative model. Similarly, for models retaining approximately 70\% of the original volume, the intragastric pressure increases to 1.3 times the pre-operative pressure. Based on these observations, the fundic pressures ($p_o$) of the pre-operative and post-operative stomach models used in this study are summarized in Table~\ref{four_cases}.
\begin{table*}[h]
    \centering
    \caption{Parameters of the four stomach models used in the simulations}
    \label{four_cases}
    \begin{tabular}{lccc}
        \toprule
        \textbf{Cases} & \textbf{Resection End Point} & \textbf{Retained Volume (\%)} & \textbf{Fundic Pressure $p_o$ (mmHg)}\\
        \midrule
        Pre-operative stomach model & --- & 100\% &  0.075 \\
        Moderate sleeve stomach model & P1 ($6\, \text{cm}$ from the pylorus) & 70\% & 0.097 \\
        Extreme sleeve stomach model & P1 ($6\, \text{cm}$ from the pylorus) & 45\% &0.125 \\
        Moderate sleeve model with reduced motility & P2 ($2\, \text{cm}$ from the pylorus) & 70\% & 0.097 \\
        \bottomrule
    \end{tabular}
    \label{four_cases}
\end{table*}



\subsection{Flow Model}
The in-house sharp-interface immersed-boundary flow solver, ViCar3D~\cite{Mittal2008}, is employed to simulate the fluid flows inside the stomach and the interaction with the stomach wall. The stomach model is immersed into a three-dimensional Cartesian volume of domain size $15 \times 10 \times 13 cm^3$ in $x$, $y$ and $z$ directions respectively. 

This solver utilizes the finite-difference method to solve the incompressible Navier-Stokes equations. Second-order central difference schemes are employed to discretize the computational domain, which consists of a total of $300 \times 200 \times 260$ Cartesian grid points with a grid spacing of $0.5\, \text{mm}$. The immersed surfaces are represented by three-dimensional surface meshes composed of more than 15,000 triangular elements. Inflow is prescribed through a cross-section at the fundus, serving as the fundic boundary, while outflow occurs through a cross-section at the end of the duodenum, serving as the duodenal boundary. To ensure numerical stability and convergence, the simulation is performed with a time step size of $0.002\, \text{s}$. Further details of the numerical setup can be found in earlier works~\cite{kuhar2022effect}.


The flow is governed by incompressible Navier–Stokes equations:
\begin{equation}\label{eqn:1}
\nabla \cdot \mathbf{u} = 0, 
\end{equation}
\begin{equation}\label{eqn:2}
\rho \left[ \frac{\partial \mathbf{u}}{\partial t} + \nabla \cdot (\mathbf{u}\mathbf{u})   \right]  = -\nabla p + \mu \nabla^2 \mathbf{u} + \rho \mathbf{g}
\end{equation}
where $\mathbf{u}$ is the fluid velocity vector, $p$ is the pressure, and $\rho$ and $\mu$ are density and dynamic viscosity respectively; $\mathbf{g}$ represents the gravity acceleration with a value of $\mathbf{g}=-9.81m/s^2$. The volume of fluid (VOF) method has been used for the multiphase flows simulations. We have two fluids in the model, one corresponds to the bolus and another that fills the stomach (this fluid has the properties of water in the current model). A single set of momentum conservation equations is used to describe the fluid as a whole:
%
\begin{equation}\label{eqn:3}
\rho_m \left[ \frac{\partial \mathbf{u}}{\partial t} + \nabla \cdot (\mathbf{u}\mathbf{u})   \right]  = -\nabla p + \mu_m \nabla^2 \mathbf{u} + \rho_m \mathbf{g}
\end{equation}
where $\rho_m$ is the mixture density and $\mu_m$ represents the mixture dynamics viscosity. The properties of the mixture of the two fluids are given by $\rho_m = \alpha_1 \rho_1 + (1 - \alpha_1) \rho_2$ and $\mu_{m} = \alpha_1 \mu_1 + (1 - \alpha_1) \mu_2$. $\alpha$ is the volume fraction, where $\alpha_1$ is the volume fraction of phase 1 and $\alpha_2$ is the volume fraction of phase 2. In the two-phase flow simulation, the volume fraction needs to satisfy $\alpha_1 + \alpha_2 = 1$. The volume fraction is solved by the advection-diffusion equation, where $D_i$ is the diffusion coefficient of phase $i$:

\begin{equation}\label{eqn:4}
\frac{\partial \alpha_i}{\partial t} + \mathbf{u} \cdot \nabla \alpha_i = D_i {\nabla}^2 \alpha_i
\end{equation}
 


\subsection{Quantities of Interest}
We are interested in gastric emptying rate and mixing inside the stomach. The gastric emptying rate quantifies how quickly stomach contents are transferred to the duodenum. To calculate this rate, we consider a cross-sectional area at the pylorus and compute the fluid flux passing through it using the following formula:
\begin{equation}\label{eqn:4}
Q_{\text{emptying}} = \int c \, \mathbf{u} \cdot \mathbf{n} \, dA,
\end{equation}
where \( c \) is the concentration of the emptied liquid, \( \mathbf{u} \) is the fluid velocity through the pylorus, \( \mathbf{n} \) is the unit normal vector of the cross-section pointing outside the stomach, and \( dA \) is the differential area element of the cross-section.

Effective digestion in the stomach relies heavily on the thorough mixing of food with gastric juice, a process that can be quantitatively analyzed using several metrics. One metric that we have used before is the Intensity of Segregation ($IoS$), which assesses mixing efficiency in binary mixtures, as detailed in Eq.~\ref{ios}:
\begin{equation}\label{ios}
IoS_i = \sqrt{\frac{\iint (C_i(\vec{x}) - \bar{C})^2 \, dA}{\iint dA}}
\end{equation}
where \( C \) denotes the concentration of either phase 1 or phase 2. $IoS$ can be directly calculated from the concentration fields of the substances involved. This metric evaluates the uniformity of a phase's distribution within the domain.  

\section{Results and Discussions} 
We first present the results on the transport of the food bolus, and we then compare the gastric emptying rates among various cases, and analyze the underlying mechanisms. Additionally, we explore the mixing of gastric contents in cases with identical sleeve sizes but different motility patterns, and discuss the relationship between mixing rates and emptying rates. The discussion of emptying rate is divided into two parts: first, we compare cases with different sleeve sizes; second, we compare cases with different motility patterns.

To validate our computational predictions, we have previously conducted a comparative analyses against established \emph{in vivo} studies from the literature. Under comparable conditions of meal properties and motility parameters, our model predicted a gastric emptying rate of $4.48 \text{ml}/\text{min}$~\cite{kuhar2022effect}. This value demonstrates good agreement with experimental measurements reported by Marciani \emph{et al.} ($4.1 \text{ml}/\text{min}$)~\cite{padalino2020effects} and Brener \emph{et al.} ($4.26 \text{ml}/\text{min}$)~\cite{brener1983regulation}. Additionally, the transpyloric pressure gradient predicted by our model for high-viscosity meals is $0.25 \text{mmHg}~ \text{cm}^{-1}$ ~\cite{kuhar2024silico}, falling within the physiological range documented in the \emph{in vivo} investigation by Indireshkumar \emph{et al.}, who reported values of $0.3 \pm 0.2 \text{mmHg}~ \text{cm}^{-1}$~\cite{indireshkumar2000relative}.

\subsection{Transportation of the Food Bolus}
\begin{figure*}[t] 
    \centering 
    \includegraphics[width=0.8\textwidth]{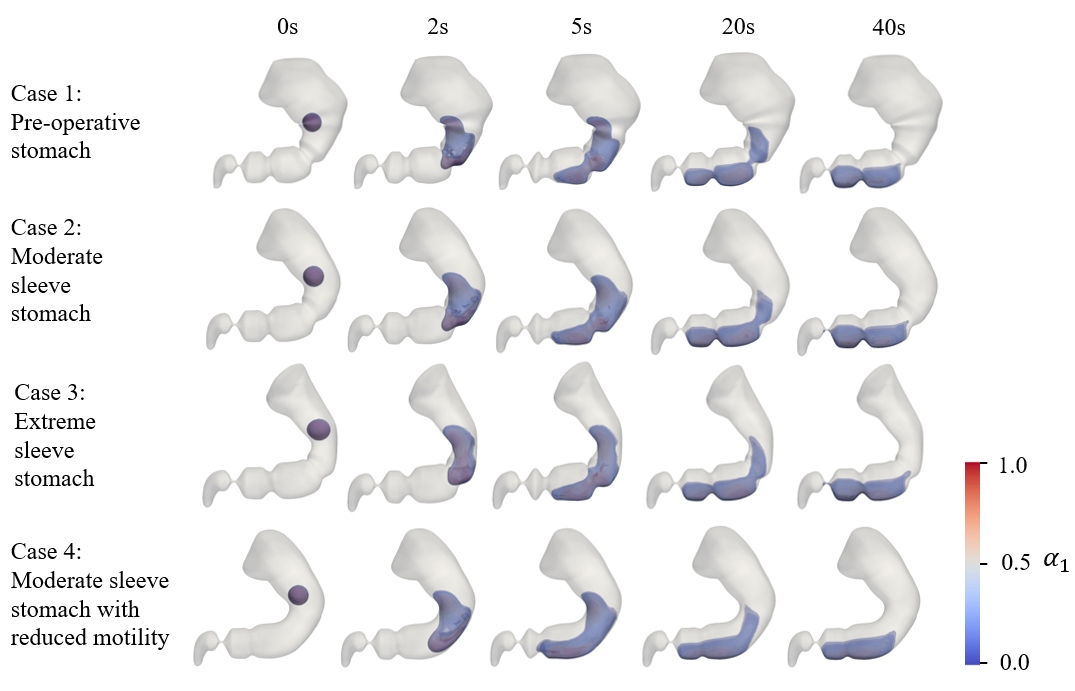} 
    \caption{Volume fraction contours of bolus in four different scenarios. $\alpha_1$ represents the volume fraction of phase 1.} 
    \label{transportation_1} 
\end{figure*}

Transportation is an important function of the stomach. After swallowing, food needs to be transported into the antrum for digestion. Thus, it plays a crucial role in regulating pH levels and digestion inside the stomach, profoundly affecting gastric functions~\cite{padalino2020effects}. The volume fraction contours of the food bolus in different scenarios are shown in Fig.~\ref{transportation_1}, illustrating how the volume fraction of the liquid meal evolves over time.

The volume fraction contours for the four cases are presented in Fig.~\ref{transportation_1}. The images illustrate the dynamic process of food bolus transport from the fundus to the antrum under the influence of gravity. After approximately 4 seconds, the bolus first reaches the antrum, marking the onset of digestion. This sequence of images clearly shows the gradual filling pattern in different regions of the stomach, where the bolus initially accumulates in the fundus and then flows downward under gravity. This time-dependent distribution provides insight into the gastric emptying function and the movement characteristics of the bolus across various gastric regions. Additionally, the color gradient in the images represents changes in bolus concentration, highlighting interactions with gastric fluids that enhance mechanical digestion. The geometry of the stomach plays a critical role in bolus flow, with its curved structure and path from the fundus to the antrum facilitating gravity-driven transport. Once deposited in the antrum, the bolus has an increased contact area with the stronger muscular contractions in this region, which accelerates mechanical breakdown and prepares the bolus for further digestion.

%
\subsection{Effects of Gastric Sleeve Size}
\begin{figure*}[t]
    \centering
    \begin{subfigure}[b]{0.32\textwidth} 
        \centering
        \includegraphics[height = 4.7cm]{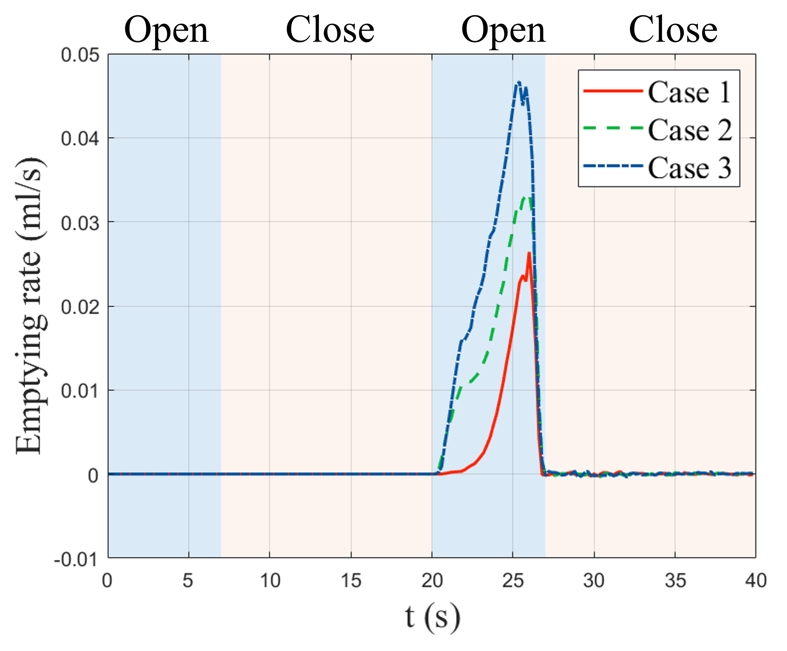} 
        \caption{Food bolus emptying rate of the pre-operative stomach, moderate sleeve stomach, and extreme sleeve stomach models.}
        \label{Scalar_1}
    \end{subfigure}
    \hfill
    \begin{subfigure}[b]{0.32\textwidth}
        \centering
        \includegraphics[height = 4.65cm]{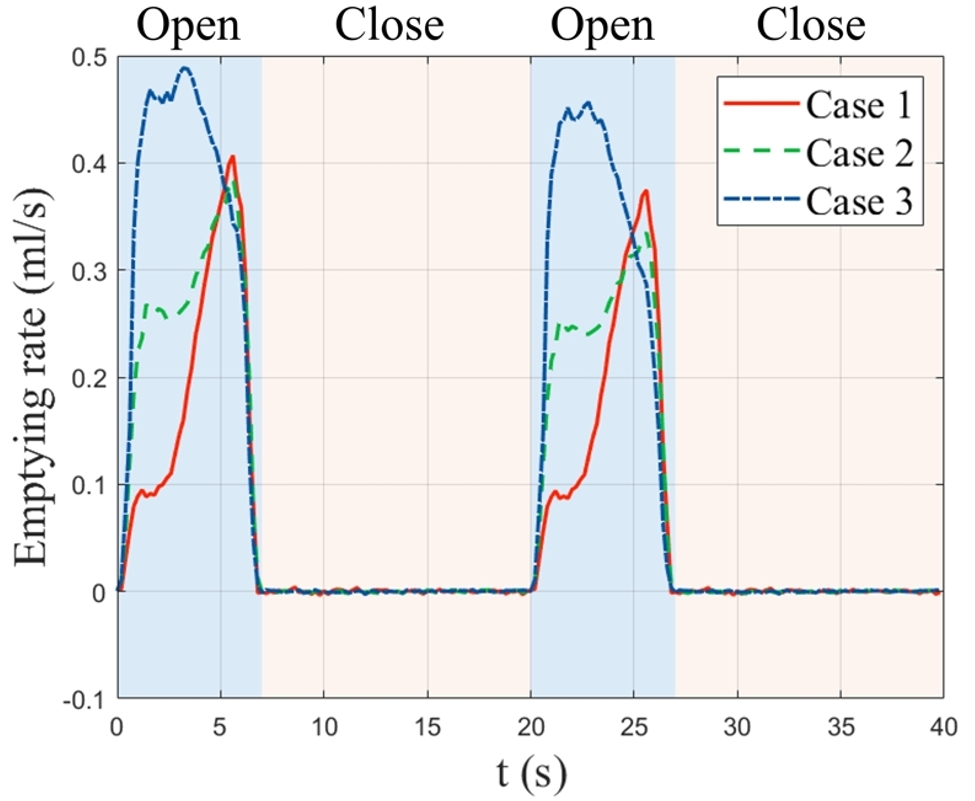} 
        \caption{Water emptying rate of the pre-operative stomach, moderate sleeve stomach, and extreme sleeve stomach models.}
        \label{Water_1}
    \end{subfigure}
    \hfill
    \begin{subfigure}[b]{0.32\textwidth} 
        \centering
        \includegraphics[height = 4.7cm]{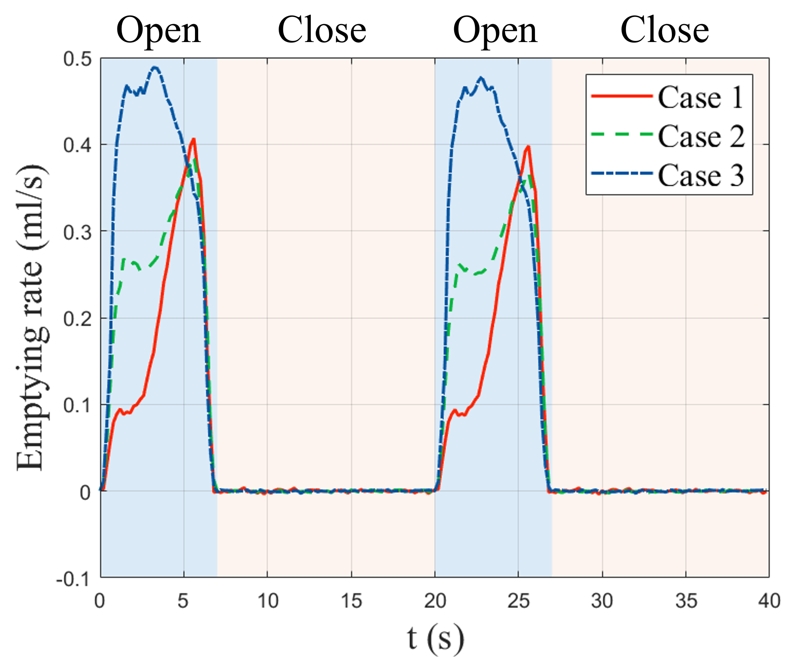} 
        \caption{Total liquid emptying rate of the pre-operative stomach, moderate sleeve stomach, and extreme sleeve stomach models.}
        \label{Total_1}
    \end{subfigure}
    \caption{Temporal evolution of emptying rates for food bolus, water, and total gastric liquid for different scenarios. Case 1: pre-operative stomach model; case 2: moderate sleeve stomach model; case 3: extreme sleeve stomach model.}
    \label{ER_1}
\end{figure*}

The gastric emptying rates for the pre-operative stomach, moderate sleeve stomach, and extreme sleeve stomach models are presented in Fig.~\ref{ER_1}. In these simulations, the motility pattern is kept similar while the sleeve size varies. From Fig.~\ref{ER_1}, it is evident that the emptying rates generally increase following laparoscopic sleeve gastrectomy. Notably, the bolus flux begins to empty during the second period, as it requires time to be transported to the antrum. The average emptying rates are calculated and presented in Table~\ref{Er_1}. For bolus flux, the average emptying rate is calculated over one antral wave cycle that extends over about 20 seconds.

In the moderate sleeve stomach model, the total gastric liquid emptying rate is 33\% faster than that of the pre-operative stomach, while in the extreme sleeve stomach model, it is 87\% faster. This indicates that a reduction in postoperative sleeve size is accompanied with a higher emptying rate. As shown in Fig.~\ref{Scalar_1} and Fig.~\ref{Water_1}, water constitutes the vast majority of the emptied liquid, exceeding 90\% of the total volume emptied. Regarding the food bolus, the emptying rate in the moderate sleeve stomach is 106\% faster than that of the pre-operative stomach, and in the extreme sleeve stomach, it is 210\% faster, indicating that liquid food empties more rapidly after surgery.

\begin{table}[h]
\caption{Average Gastric Emptying Rates (ml/s) for Stomach Models with Different Sleeve Sizes
\label{Er_1}}
\setlength{\tabcolsep}{4pt}
\centering{%
\begin{tabular}{l c c c c c}
\toprule
\makecell{Averaged \\ emptying rate \\ (ml/s)} & Pre-operative & Moderate sleeve &  Extreme sleeve \\
\midrule
Bolus flux  & $0.0073$ & $0.0155$ & $0.0226$ \\
Water flux  & $0.1786$ & $0.2343$ & $0.3295$ \\
Total flux  & $0.1822$ & $0.2421$ & $0.3408$ \\

\end{tabular}
}%
\end{table}

\begin{figure*}[h] 
    \centering
    \begin{subfigure}[b]{0.46\textwidth} 
        \centering
        \includegraphics[height = 7cm]{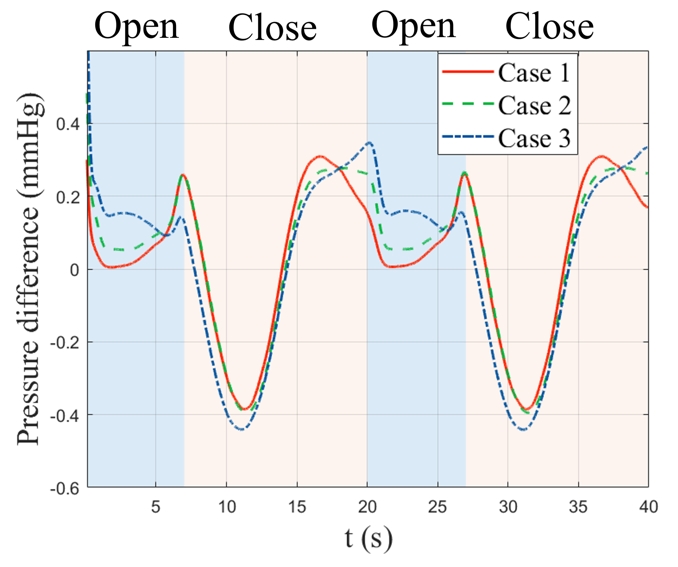} 
        \caption{Intragastric pressure evolution for the pre-operative stomach, moderate sleeve stomach, and extreme sleeve stomach models.}
        \label{p_1}
    \end{subfigure}
    \hfill
    \begin{subfigure}[b]{0.46\textwidth}
        \centering
        \includegraphics[height = 7cm]{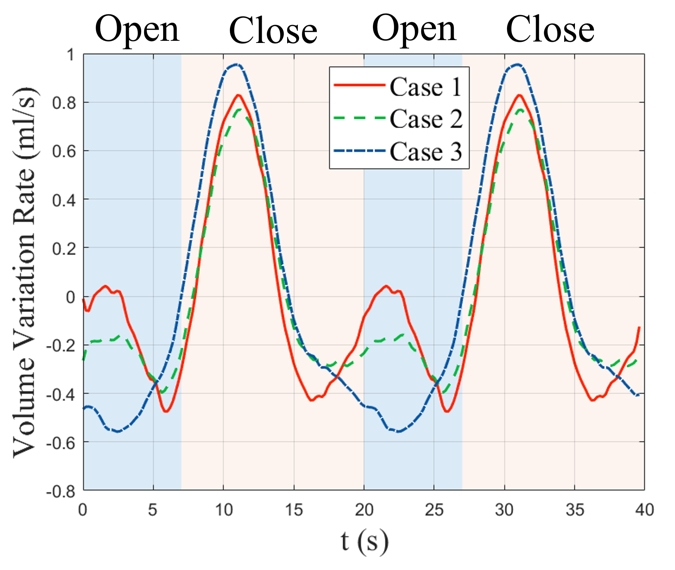} 
        \caption{Volume variation rate evolution for the pre-operative stomach, moderate sleeve stomach, and extreme sleeve stomach models.}
        \label{v_1}
    \end{subfigure}
    \hfill
    \caption{Intragastric pressure and volume variation rate evolution for different scenarios. Case 1: pre-operative stomach model; case 2: moderate sleeve stomach model; case 3: extreme sleeve stomach model. The pressure at the duodenal outlet is set to zero Pascal.}
    \label{pressure_volume}
\end{figure*}

Gastric emptying rate can be affected by multiple factors, with intragastric pressure and volume variation playing a significant role. Fig.~\ref{pressure_volume} shows the intragastric pressure and volume variation rates in pre-operative, moderate sleeve and extreme sleeve stomach models. 
Fig.~\ref{p_1} shows the temporal intragastric pressure evolution for the pre-operative stomach, moderate sleeve stomach, and extreme sleeve stomach models. In our simulations, the pressure at the duodenal outlet is set to a reference value of zero, serving as the baseline. A pressure value greater than zero indicates that the local pressure is higher than at the outlet, while a value less than zero indicates a lower pressure than at the outlet. It is observed that during the opening period of pylorus, the intragastric pressure in the extreme sleeve stomach is evidently higher than in the pre-operative stomach, and the pressure in the moderate sleeve stomach is also slightly higher. The average intragastric pressure during the periods when the pylorus is open is calculated and presented in Table~\ref{pressure_1}. This suggests that the reduced gastric volume resulting from LSG leads to increased intragastric pressure during pylorus opening, which contributes to a faster gastric emptying rate.
\begin{table}[h]
\caption{Average Intragastric Pressure (mmHg) for Stomach Models with Different Sleeve Sizes}
\label{pressure_1}
\setlength{\tabcolsep}{4pt}
\centering
\begin{tabular}{l c c c}
\toprule
& Pre-operative & Moderate sleeve & Extreme sleeve \\
\midrule
\makecell{Averaged \\ intragastric \\ pressure \\ (mmHg)} & $0.1745$ & $0.3192$ & $0.4432$ \\
\bottomrule
\end{tabular}
\end{table}

Fig.~\ref{v_1} shows the rates of volume variation for the pre-operative stomach, moderate sleeve stomach, and extreme sleeve stomach models. Positive values indicate an increase in volume, while negative values indicate a decrease. During the pylorus opening period, the stomach contracts and ACWs propagate along the antrum towards the pylorus, with the terminal antral contraction (TAC) causing a decrease in volume. When the pylorus is closed, the muscles in the antrum relax, and the volume increases to the original value. Both ACWs and TAC significantly affect the stomach volume variation, and thus the gastric emptying rate. During the pylorus opening period, the gastric volume variation rate is negative, with greater magnitudes observed in the moderate sleeve and extreme sleeve stomach models compared to the pre-operative stomach model, indicating a faster rate of volume decrease. When comparing the moderate sleeve with the extreme sleeve stomach, the volume decrease is even more pronounced in the extreme sleeve model. This increased rate of volume change, combined with elevated intragastric pressure, accounts for the acceleration of gastric emptying following laparoscopic sleeve gastrectomy. 

The increase in the volume variation rate after LSG can be attributed to changes in the propagation of antral contraction waves. In a normal stomach, ACWs originate from the proximal stomach (with the starting point at \( l_A = 8\, \text{cm} \) from the pylorus) and gradually increase in amplitude as they propagate toward the pylorus, reaching their maximal value in the antrum. After LSG, due to impairment of the pacemaker in the corpus, the ACWs begin to grow from the distal stomach (with the starting point in our model at \( l_A = 6\, \text{cm} \) from the pylorus for the moderate sleeve stomach and \( l_A = 2\, \text{cm} \) from the pylorus for the extreme sleeve stomach) and reach their maximal amplitude more quickly. Consequently, the stomach volume variation occurs more rapidly because the ACWs attain their maximum amplitudes sooner. This explains why the greater the volume resected during LSG, the faster the volume variation rate. The volume reduced after laparoscopic sleeve gastrectomy impairs the storage function of the stomach and alters the intragastric pressure during digestion of the same food volume, profoundly influencing the gastric emptying rate.

\subsection{Effects of Gastric Motility}
\begin{figure*}[t]
    \centering
    \begin{subfigure}[b]{0.32\textwidth} 
        \centering
        \includegraphics[height = 4.6cm]{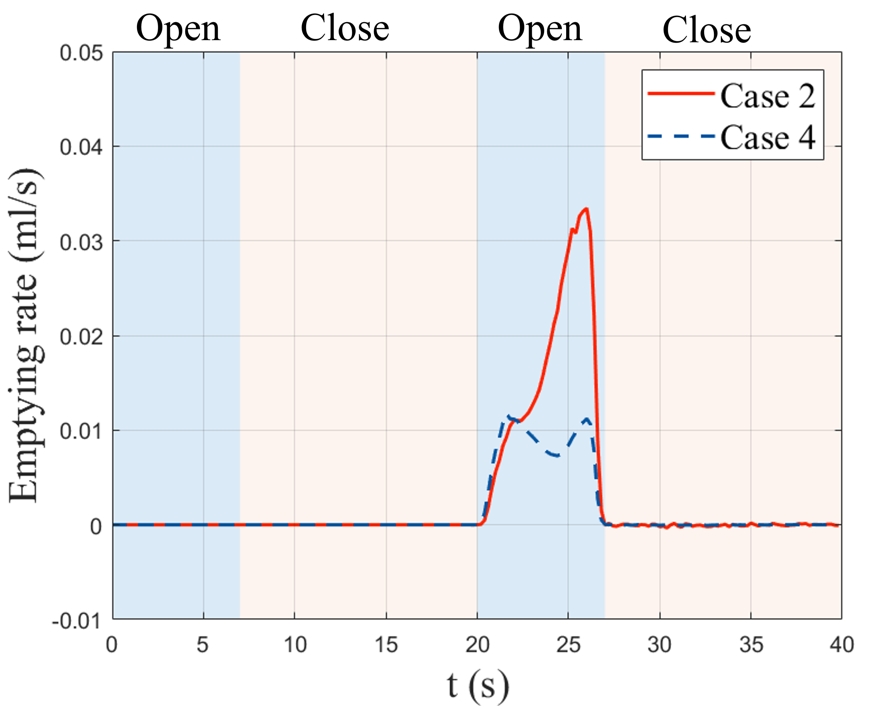} 
        \caption{Bolus emptying rate of the moderate sleeve stomach, and moderate sleeve stomach with reduced motility models.}
        \label{Scalar_2}
    \end{subfigure}
    \hfill
    \begin{subfigure}[b]{0.32\textwidth}
        \centering
        \includegraphics[height = 4.6cm]{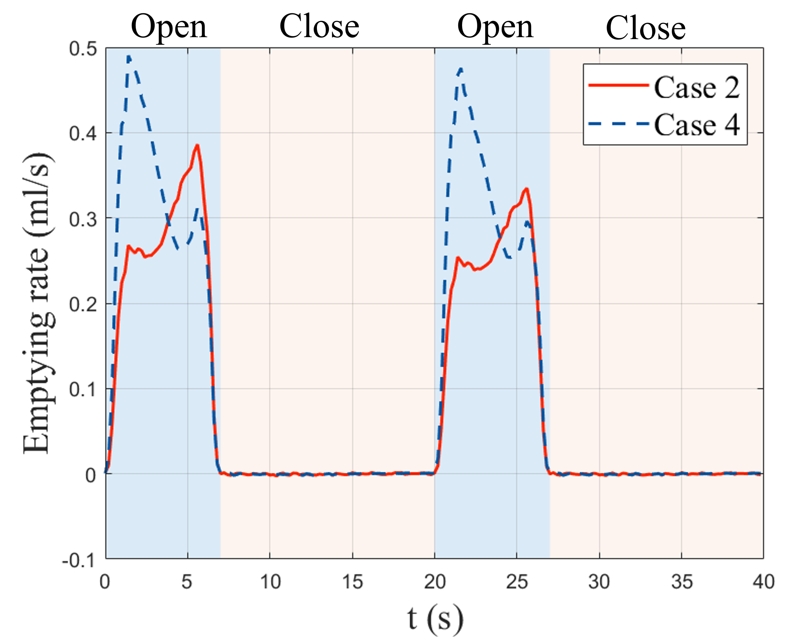} 
        \caption{Water emptying rate of the moderate sleeve stomach, and moderate sleeve stomach with reduced motility models.}
        \label{Water_2}
    \end{subfigure}
    \hfill
    \begin{subfigure}[b]{0.32\textwidth} 
        \centering
        \includegraphics[height = 4.6cm]{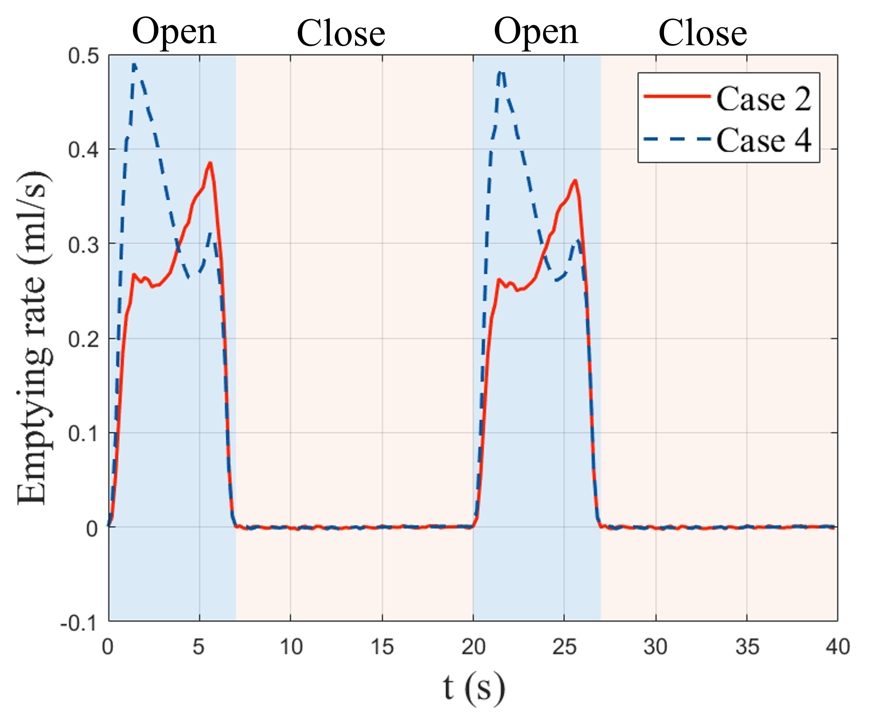} 
        \caption{Total gastric liquid emptying rate of the moderate sleeve stomach, and moderate sleeve stomach with reduced motility models.}
        \label{Total_2}
    \end{subfigure}
    \caption{Temporal evolution of emptying rates for bolus, water, and total gastric liquid for different scenarios. Case 2: moderate sleeve stomach model; case 4: moderate sleeve stomach with reduced motility model.}
    \label{ER_2}
\end{figure*}

The gastric emptying rates for stomach models with different motility patterns are presented in Fig.~\ref{ER_2}. The reduced motility model exhibits a 21\% faster emptying rate of total gastric liquid compared to the moderate sleeve stomach model, while the emptying rate of food bolus is 49\% slower. 
\begin{table}[h]
\caption{Average Gastric Emptying Rates (ml/s) for Stomach Models with Different Motility Patterns}
\label{Er_2}
\setlength{\tabcolsep}{4pt}
\centering
\begin{tabular}{l c c}
\toprule
\makecell{Averaged \\ emptying rate \\ (ml/s)} & Moderate sleeve & \makecell{Moderate sleeve with \\ reduced motility} \\
\midrule
Bolus flux  & $0.0155$ & $0.0079$ \\
Water flux  & $0.2343$ & $0.2894$ \\
Total flux  & $0.2421$ & $0.2933$ \\
\bottomrule
\end{tabular}
\end{table}

\begin{figure*}[h] 
    \centering
    \begin{subfigure}[b]{0.46\textwidth} 
        \centering
        \includegraphics[height = 7cm]{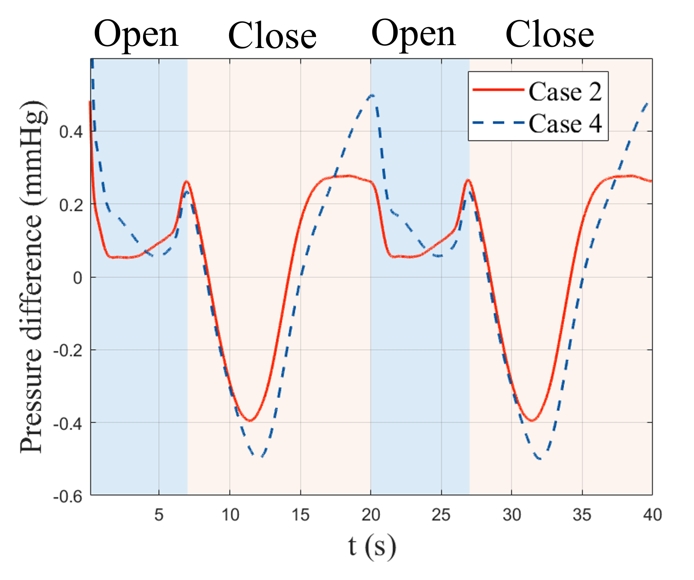} 
        \caption{Intragastric pressure evolution for the pre-operative stomach, moderate sleeve stomach, and extreme sleeve stomach models.}
        \label{p_2}
    \end{subfigure}
    \hfill
    \begin{subfigure}[b]{0.46\textwidth}
        \centering
        \includegraphics[height = 7cm]{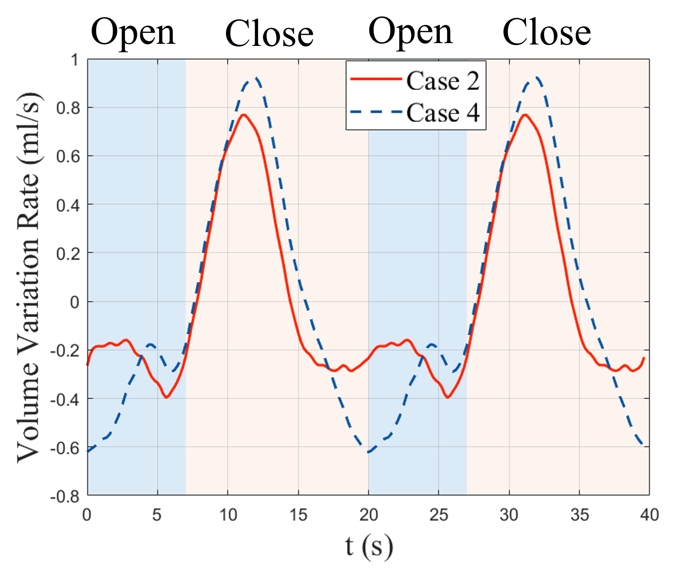} 
        \caption{Volume variation rate evolution for the pre-operative stomach, moderate sleeve stomach, and extreme sleeve stomach models.}
        \label{v_2}
    \end{subfigure}
    \hfill
    \caption{Intragastric pressure and volume variation rate evolution for different scenarios. Case 2: moderate sleeve stomach model; case 4: moderate sleeve stomach with reduced motility model. The pressure at the duodenal outlet is set to zero.}
    \label{pressure_volume_2}
\end{figure*}

The temporal intragastric pressure evolutions for moderate sleeve stomach models with nominal motility (Case-2) and reduced motility (Case-4) are shown in Fig.~\ref{p_2}. During the pylorus opening periods, the average pressure is higher in the reduced motility stomach, corresponding to its faster emptying rates. The pressure evolution profiles in these two cases differ in shape, leading to varying emptying rate evolutions among the scenarios. The emptying rate profiles are positively correlated with the corresponding intragastric pressure evolution profiles. Specifically, in the moderate sleeve stomach model, the pressure is smaller in the early phase and larger in the later phase, resulting in a peak of emptying rate towards the end of the opening period, just before the pylorus closes. The reduced motility stomach exhibits the opposite situation, with higher pressure in the early time and lower pressure later on. This indicates that intragastric pressure has a profound influence on the gastric emptying rate. 
Fig.~\ref{v_2} illustrates the volume variation rates for the two cases. During the pylorus opening period, the total volume decrease is greater in the reduced motility stomach model and smaller in the moderate sleeve stomach model. In the reduced motility case, the point at which the ACWs start to grow is even closer to the pylorus (with the starting point at \( l_A = 2\, \text{cm} \) compared to \( l_A = 6\, \text{cm} \) in the moderate sleeve stomach model). Consequently, the ACWs in the reduced motility stomach model reach their maximal amplitude more quickly, leading to a higher volume variation rate. This corresponds to the gastric emptying rates, which are faster in the reduced motility stomach model and slower in the moderate sleeve stomach model.


Several studies utilizing experimental and clinical data have also found that gastric emptying significantly accelerates after LSG. Specifically, Sista \emph{et al.}~\cite{sista2017effect} observed a significant acceleration in postoperative gastric emptying, with the half-emptying times (\( T^{1/2} \)) for liquids and solids reduced from 26.7 minutes and 68.7 minutes preoperatively to 15.2 minutes and 33.5 minutes postoperatively, respectively. Similarly, Kara \emph{et al.}~\cite{kara54} reported that the preoperative and postoperative liquid-phase gastric emptying half-times were 41.86 minutes and 6.82 minutes, respectively. These findings align with our results.

Furthermore, Kara \emph{et al.}~\cite{kara54} found that rapid gastric emptying supports weight loss based on their experimental data. Research indicates that the underlying mechanisms by which rapid gastric emptying promotes weight loss are related to hormonal responses and reduced nutrient absorption. Accelerated emptying increases the release of hormones such as glucagon-like peptide-1 (GLP-1), which enhance satiety and reduce appetite, thereby helping patients consume less food. This hormonal change contributes to sustained weight reduction after surgery. Additionally, faster gastric emptying decreases the residence time of food in the stomach, reducing the effectiveness of digestion and calorie absorption, which further contributes to weight loss~\cite{sista2017effect, kara54, garay2018influence}.

\begin{figure}[t] 
    \centering 
    \includegraphics[width=0.46\textwidth]{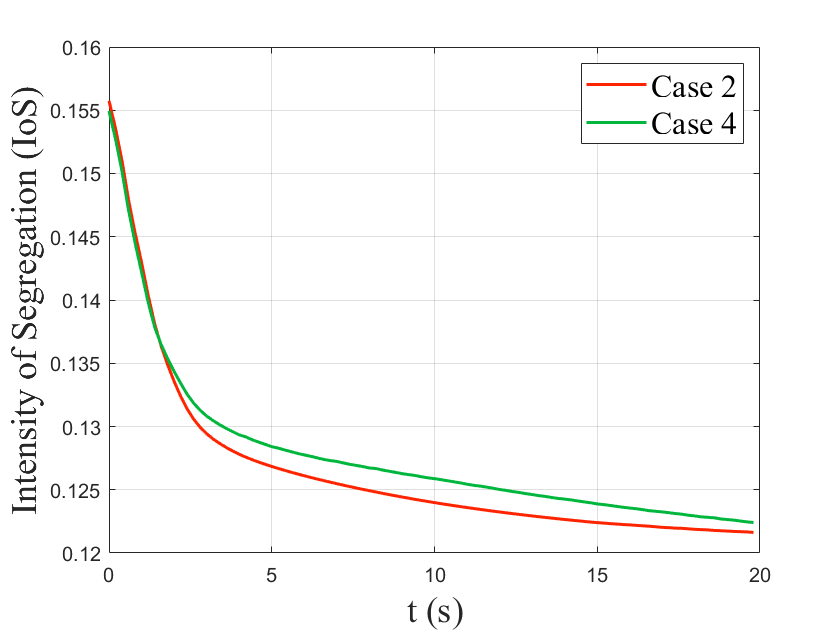} 
    \caption{Gastric mixing rate of scenarios with different motility patterns. Case 2: moderate sleeve stomach model; case 4: moderate sleeve stomach with reduced motility model.} 
    \label{mixing} 
\end{figure}

Fig.~\ref{mixing} illustrates the mixing conditions during the first period ($0 \sim 20s$) before the onset of bolus emptying. The mixing is quantified by $IoS$ of phase 1 (the liquid meal) in the reduced motility model and moderate sleeve stomach model. Initially, the $IoS$ values for the two scenarios are identical, reflecting identical initial conditions. As time progresses, the liquid meal begins to move towards the antrum and simultaneously disperses in the gastric solvent, leading to a rapid decline in $IoS$. Since $IoS$ measures the degree of separation, a lower $IoS$ value indicates improved mixing. Fig.~\ref{mixing} clearly shows that the reduced motility model exhibits the slower mixing. This suggests that reduced gastric motility negatively impacts the mixing efficiency, and consequently, food digestion within the stomach. 

\begin{figure*}[h] 
    \centering 
    \includegraphics[width=0.8\textwidth]{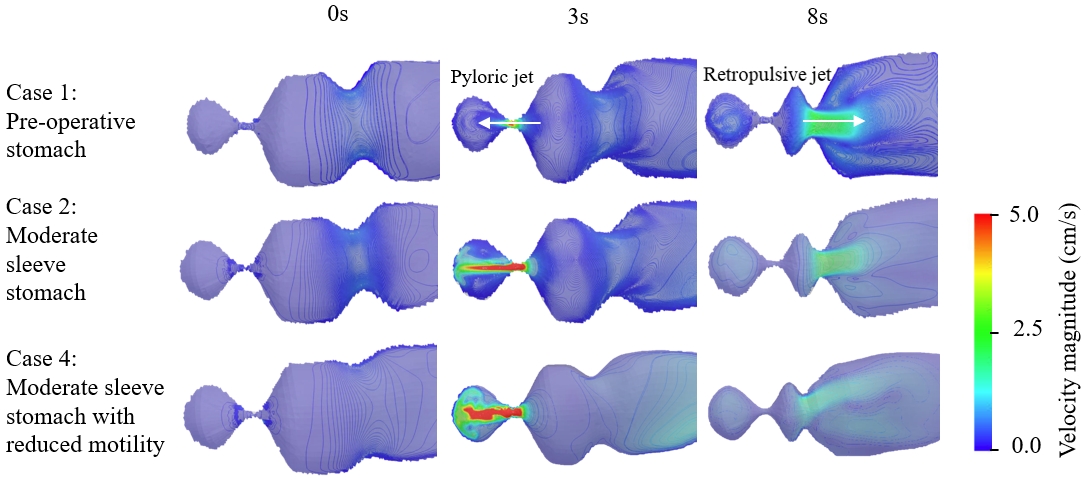} 
    \caption{Velocity magnitude contours for stomach models with varying motility are shown on a cross-section through the antro-duodenal region at three time points: t=0, t=3 and t=8.} 
    \label{contour} 
\end{figure*}

Fig.~\ref{contour} illustrates the velocity magnitude contours across the antro-duodenal region for stomach models with varying motility parameters, observed at three distinct temporal points. At $t=0$, the pylorus exhibits complete closure while an ACW approaches the pyloric region. At $t=3s$, with the pylorus in its open configuration, a characteristic pyloric jet emerges, facilitating the emptying of gastric contents from the antrum to the duodenum. These contours reveal enhanced pyloric jet velocity in the post-operative stomach models, correlating with accelerated gastric liquid emptying rates observed post-operatively. At $t=5s$, immediately following pyloric closure, the TAC persists, generating a significant retrograde flow pattern. This retrograde jet plays a crucial role in gastric mixing dynamics, where increased jet velocity significantly facilitates antral mixing. The figures demonstrate attenuated retrograde jet formation in post-operative models. Furthermore, comparative analysis between post-operative cases reveals further diminished retrograde jet intensity in the reduced motility model, explaining the observed lower mixing efficiency in this model.
%

We further recall the food bolus and total gastric liquid emptying rates of these two models here, shown in Fig.~\ref{Scalar_2} and Fig.~\ref{Total_2}. It is evident that this reduction in mixing efficiency influences the emptying rate of the food bolus. Although the total gastric liquid emptying rate is higher in the reduced motility stomach model, the food bolus emptying rate is only half of that of the moderate sleeve stomach model due to slower mixing before emptying.


\section{Conclusions}
In this study, we utilized multiphase flows simulations to investigate the effects of LSG on gastric digestion and emptying functions. By constructing several models of pre- and post-operative stomachs, we analyzed how changes in stomach size and gastric motility due to LSG impact the emptying and mixing of gastric contents.

Our simulations demonstrated that the reduction in sleeve size and the concomitant increase in fundic pressure following LSG leads to a significant increase in the gastric emptying rate. The total gastric liquid emptying rate increased by 33\% in the moderate sleeve stomach model and by 87\% in the extreme sleeve stomach model compared to the pre-operative stomach. The emptying rate of the food bolus was 106\% faster in the moderate sleeve stomach and 210\% faster in the extreme sleeve stomach, indicating a substantial acceleration of liquid food emptying after surgery.

Several \emph{in vivo} studies have similarly reported significant acceleration in gastric emptying following LSG, with reductions in half-emptying times ($T_{1/2}$) for liquids and solids. For example, Sista \emph{et al.}~\cite{sista2017effect} observed that postoperative $T_{1/2}$ decreased from 26.7 minutes to 15.2 minutes for liquids, and from 68.7 minutes to 33.5 minutes for solids. Similarly, Kara \emph{et al.}~\cite{kara54} reported reductions in liquid-phase gastric emptying $T_{1/2}$ from 41.86 minutes preoperatively to 6.82 minutes postoperatively. These findings align with our simulation results. Rapid gastric emptying is thought to promote weight loss by enhancing hormonal responses—such as increased release of glucagon-like peptide-1 (GLP-1), which enhances satiety and reduces appetite—and by decreasing nutrient absorption due to reduced digestion time~\cite{sista2017effect, kara54, garay2018influence}. 

Intragastric pressure emerged as a key factor affecting the gastric emptying rate. Our results showed that reduced sleeve size and altered motility patterns lead to changes in intragastric pressure profiles. Higher intragastric pressures were associated with increased emptying rates. Specifically, during the pylorus opening period, the extreme sleeve stomach exhibited evidently higher intragastric pressure compared to the pre-operative stomach, correlating with its faster emptying rate. The pressure evolution profiles differed among the models, leading to varying emptying rates over time. 

The volume variation rates further supported these findings. When the pylorus was open, the decrease in volume was more pronounced in the post-surgery stomach models. The differences in emptying rates among the various models were attributed to the combined effects of altered intragastric pressure and volume variation rates.

The study also revealed that the mixing rate was notably influenced by gastric motility patterns. The moderate sleeve stomach model exhibited better mixing performance, while the corresponding model with reduced motility showed decreased mixing rates. Impaired mixing can slow down the digestion process and influence the emptying rate of nutrients.

While this study provides significant insights into the biomechanical effects of laparoscopic sleeve gastrectomy on gastric digestion and emptying functions, certain limitations should be noted. Firstly, our simulations considered only liquid meals, which may not fully capture the complexities involved in the digestion of solid or semi-solid foods. Secondly, we employed a simplified model of postoperative gastric motility, assuming a complete loss of peristalsis in the sleeve tube with preserved antral motility. In reality, gastric motility patterns after LSG can vary among patients and may not be entirely absent in the sleeve portion. Despite these limitations, this work offers novel data and valuable insights into the digestive processes following LSG. Moreover, it represents a novel effort in applying computational fluid dynamics to study the effects of LSG, highlighting the potential of computational models to enhance surgical planning and postoperative management. This foundational work also paves the way for future studies to develop more complex models and explore a wider range of meal textures and motility conditions.

Future research could focus on incorporating more detailed gastric motility patterns, including patient-specific variations, to make the model more accurate. Investigating the effects of solid meals and varying meal viscosities could provide a more comprehensive understanding of post-surgery digestion. Additionally, integrating hormonal and neural feedback mechanisms into the models may offer deeper insights into the regulatory processes affecting gastric functions after LSG. Clinical studies correlating computational findings with patient outcomes would further validate the models and support the development of personalized surgical approaches and postoperative management strategies.

Overall, this study provides significant and comprehensive insights into the biomechanical effects of laparoscopic sleeve gastrectomy on gastric functions. Our findings demonstrate that both the reduction in sleeve size and the impairment of gastric motility significantly alter the stomach's emptying and mixing processes, which are closely linked to the weight loss outcomes and nutritional status of patients undergoing LSG. By the application of computational fluid dynamics to model these complex physiological changes, we have not only advanced the understanding of gastric mechanics post-surgery but also highlighted the potential of computational models to inform surgical planning and improve patient outcomes. This study establishes the use of computational fluid dynamics in the study of bariatric surgery, paving the way for future research to develop more detailed models and further enhance surgical planning and patient care.




\section*{Acknowledgment} 
The authors gratefully acknowledge the financial support provided by the National Science Foundation via Award no. CBET
2019405. We also extend our thanks to the Advanced Research Computing at Hopkins (ARCH) for granting access to their computational resources, which were essential for performing the simulations presented in this study.




\appendix   
\section{Validation of the VOF Method -- Two-Dimensional Falling Drop Simulation}\label{app}
\begin{figure}[h] 
\centering \includegraphics[width=0.46\textwidth]{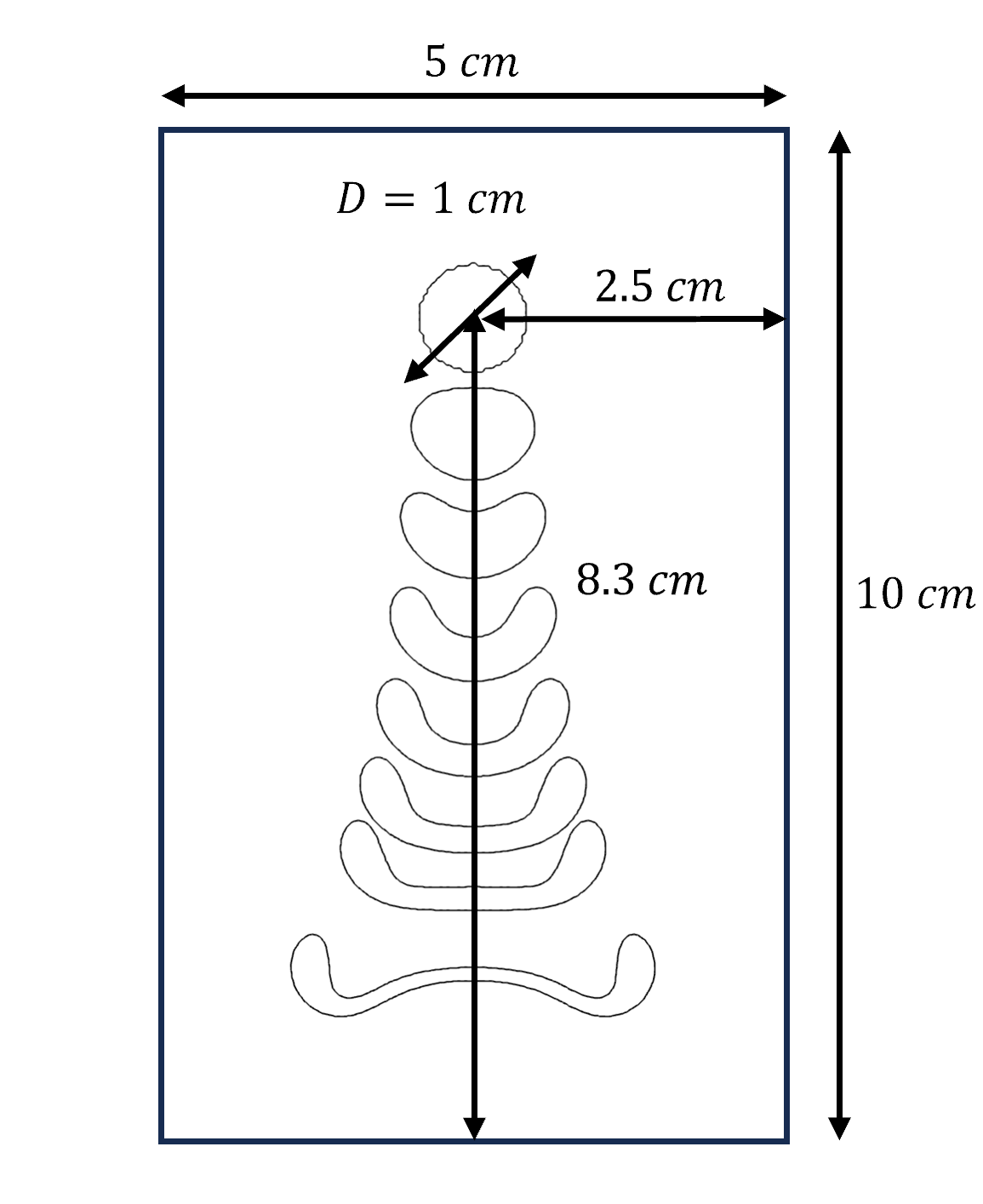} 
\caption{Schematic configuration of the two-dimensional falling drop simulation along with drop shapes at different time steps.} \label{configuration} 
\end{figure}

\begin{figure*}[t] 
\centering 
\begin{subfigure}[b]{0.46\textwidth} \centering 
\includegraphics[height=6.5cm]{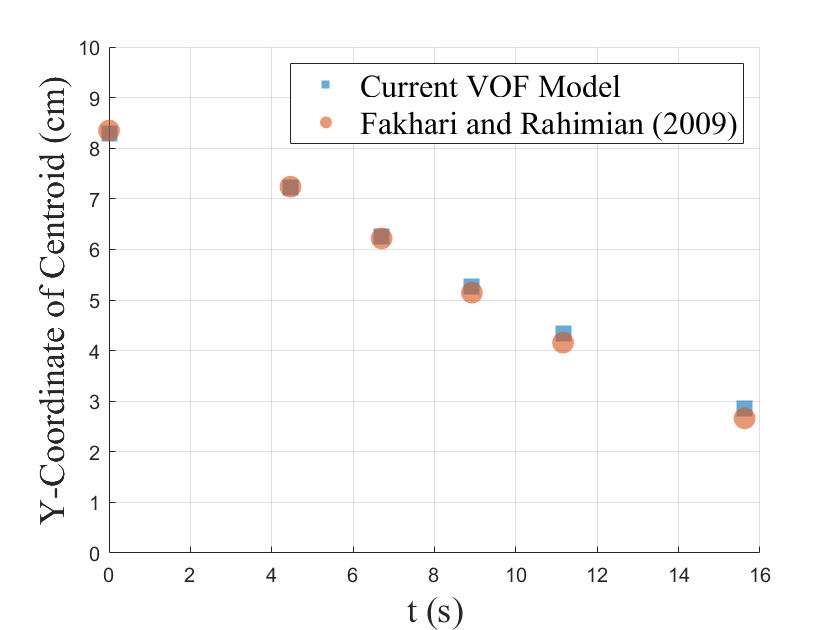} 
\caption{Evolution of y-coordinate of the drop's centroid over time.} \label{y_centroid} 
\end{subfigure}
\hfill 
\begin{subfigure}[b]{0.46\textwidth} \centering 
\includegraphics[height=6.5cm]{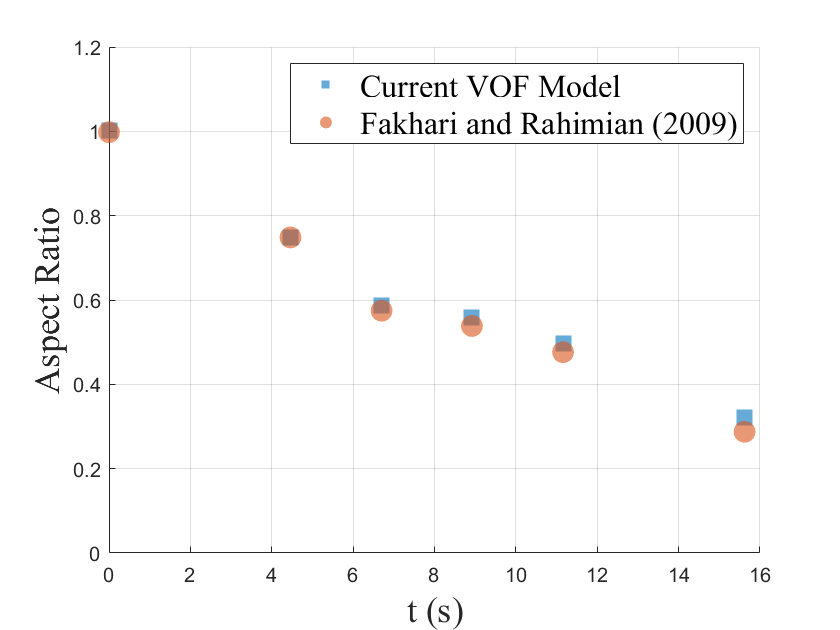} \caption{Evolution of the aspect ratio of the drop over time.} 
\label{ar} 
\end{subfigure} 
\caption{Comparison of simulation results with those of Fakhari and Rahimian~\cite{fakhari2009simulation}.} \label{app_results} 
\end{figure*}

To validate our VOF code, we conducted a two-dimensional simulation of a falling drop and compared the results with those reported by Fakhari and Rahimian~\cite{fakhari2009simulation}. The configuration of this validation case is illustrated in Fig.~\ref{configuration}, along with the drop shapes at different time steps. The drop has a diameter of $D = 1,\mathrm{cm}$. The computational domain measures $5D$ in width and $10D$ in height. The density ratio between the drop and the ambient fluid is 5, and both fluids have the same kinematic viscosity. A gravitational acceleration of $1\mathrm{cm/s}^2$ acts downward. The grid resolution is $456 \times 917$ in the $x$ and $y$ directions, respectively. The simulation results are presented in Fig.~\ref{app_results}. Fig.~\ref{y_centroid} compares the vertical position of the drop's centroid at various time steps, while Fig.~\ref{ar} shows the evolution of the aspect ratio over time. These figures demonstrate good agreement with the data from Fakhari and Rahimian~\cite{fakhari2009simulation}, indicating the accuracy and reliability of our VOF code.



\bibliographystyle{asmejour}   

\bibliography{export} 



\end{document}